\documentstyle[prl,aps]{revtex}
\newcommand{\be}{\begin{equation}}
\newcommand{\ee}{\end{equation}}
\newcommand{\ba}{\begin{eqnarray}}
\newcommand{\ea}{\end{eqnarray}}

\def\a{\alpha}

\def\c{\chi}
\def\d{\delta}

\def\vf{\varphi}

\def\j{\psi}

\def\m{\mu}
\def\n{\nu}
\def\o{\omega}

\def\x{\xi}

\def\D{\Delta}

\def\G{\Gamma}

\def\Ld{\Lambda}

\def\S{\Sigma}

\def\cf{{\cal F}}

\def\cm{{\cal M}}

\def\co{{\cal O}}

\def\cs{{\cal S}}

\def\cv{{\cal V}}

\newcommand{\ov}{\overline}
\newcommand{\ti}{\tilde}
\newcommand{\wt}{\widetilde}

\newcommand{\pa}{\partial}
\def\sl#1{\rlap{\hbox{$\mskip 1 mu /$}}#1}
\def\Sl#1{\rlap{\hbox{$\mskip 3 mu /$}}#1}

\begin{document}
\draft
\title{Probing the Nielsen identities}
\author{O.M. Del Cima\thanks{
Supported by the {\it Fonds zur F\"orderung der Wissen\-schaftlichen 
Forschung (FWF)} under the contract number P11654-PHY.}\thanks{E-mail: 
delcima@tph73.tuwien.ac.at .} }
\address{{\it Institut f\"ur Theoretische Physik,}\\
Technische Universit\"at Wien (TU-Wien),\\
Wiedner Hauptstra{\ss}e 8-10 - A-1040 - Vienna - Austria. }
\date{\today}
\maketitle
\begin{abstract}

We present some comments concerning the validity of the Nielsen 
identities for renormalizable theories quantized in general linear 
covariant gauges in a context of compact gauge Lie groups.

\end{abstract}
\pacs{PACS numbers: 11.10.Gh 11.15.-q 11.15.Bt 11.15.Ex 
\hspace{4,96cm}TUW-99-04}


The gauge dependence of the effective potential is controlled 
by a set of differential equations called  Nielsen 
identities~\cite{nielsen}, which are satisfied 
whether the gauge symmetry is spontaneously broken or not. In 
the case of spontaneously broken gauge theories, one of the physical 
appeals of the Nielsen identities is that they guarantee that
the spontaneous symmetry breaking is realized by an actual physical 
minimum, in spite of the effective potential being gauge dependent.

Physical observables gauge independence has been studied in connection 
with the idea of extended BRS symmetry~\cite{sz,piguet1,piguet2}, 
this is achieved by enlarging the BRS symmetry allowing, in such a way, 
the transformation of the gauge parameters into new (BRS invariant) 
Grassmann variables.  
  
The main purpose of this letter is to present some comments with respect to 
the Nielsen identities in a context of renormalizable theories 
quantized in general linear covariant gauges using the 
extended BRS technique. All subtleties of the method used here, 
as well as some controversies presented in the literature related 
to the validity of the Nielsen identities, are discussed quite 
detailed in~\cite{our}. 

Let us consider a general compact Lie group, $G=S\oplus A$, 
with $S$ and $A$ being the semi-simple and the Abelian factors, respectively. 
The matter content is described by a set of scalar fields, 
$\vf_i(x)$ ($i=1,\dots,n_\vf$), and spinor fields, $\j_I(x)$ 
($I=1,\dots,n_\j$). The matter fields carrying anti-Hermitian 
irreducible representations of $G$ transform as
\ba
&&\d\vf_i(x)=\o^a(x)T_{a i}^{(\vf) j}(\vf_j(x)+v_j)~,\nonumber\\
&&\d\j_I(x)=\o^a(x)T_{a I}^{(\j) J}\j_J(x)~~,\label{gtransfm}
\ea
while the transformation of gauge fields $A_\m^a(x)$ is
\be
\d A_\m^a(x)=\pa_\m\o^a(x)-f_{bc}^a\o^b(x)A_\m^c(x)~~, \label{gtransfg}
\ee
where $\o^a(x)$ are real $C^\infty$ functions of the space-time point 
$x$, and $f_{bc}^a$ are the real structure constants of $G$. The set of 
indices $a,b,c$ should be understood as a split into a semi-simple 
subset $a_S,b_S,c_S=1,\dots,N_S$ and an Abelian subset 
$a_A,b_A,c_A=1,\dots,N_A$, corresponding to the semi-simple and the Abelian 
components of $G$. 

Owing to the local character of the transformations (\ref{gtransfm}), 
covariant derivatives are required, they read
\ba
&&D_\m\vf_i(x)=\pa_\m\vf_i(x)
-A_\m^a(x)T_{a i}^{(\vf) j}\vf_j(x)~~,\nonumber\\
&&{\Sl D}\j_I(x)={\sl\pa}\j_I(x)
-{\Sl A}^a(x)T_{a I}^{(\j) J}\j_J(x)~~,\label{covdev}
\ea
with the field-strength given by
\be
F_{\m\n}^a=\pa_\m A_\n^a - \pa_\n A_\m^a - f_{bc}^a A_\m^b A_\n^c~~.
\ee

Let us consider, {\it a priori}, a renormalizable theory~\cite{piguet2} 
in $D$ space-time dimensions represented by the classical action
$\S_{\rm inv}$, build up in terms of the matter fields, 
$\vf_i$ and $\j_I$, the field-strength, $F_{\m\n}^a$, 
and the covariant derivatives (\ref{covdev}), which shall be, moreover, 
invariant under the following BRS transformations:
\ba
&&sA_\m^{a_S}=D_\m c^{a_S}\equiv(\pa_\m c^{a_S} - f_{bc}^{a_S} A_\m^b c^c)~,
~~sA_\m^{a_A}=\pa_\m c^{a_A}~~,\nonumber\\
&&sc^{a_S}=\frac12 f_{bc}^{a_S} c^b c^c~,~~sc^{a_A}=0~~,\nonumber\\
&&s\vf_i=c^a T_{a i}^{(\vf) j}(\vf_j+v_j)~,
~~s{\ov c}^a=b^a~~,\nonumber\\
&&s\j_I=c^a T_{a I}^{(\j) J}\j_J~,
~~sb^a=0~~.\label{brs}
\ea

As we have previously mentioned, we shall restrict our analysis to a 
general covariant linear gauge, which should be added to the classical 
action, $\S_{\rm inv}$, in order to make the quantization feasible. 
A general linear BRS invariant gauge-fixing~\cite{brs,bbbc} 
(due to the nilpotency of $s$) can be written as
\ba
\S_{\rm gf}&=&\int d^Dx~\Ld^{ab}\left\{b_a [g_b + {\frac{\x^0}2}b_b] - 
{\ov c}_a(\cm c)_b\right\} \nonumber\\
&=&s\int d^Dx~\Ld^{ab}{\ov c}_a\left\{g_b + 
{\frac{\x^0}2} b_b\right\}~~,
\label{gf}
\ea
where $\Ld^{ab}$ is an arbitrary symmetric positive definite matrix, 
$\x^0$ is a gauge parameter and $g_a(x)$ is the gauge function 
(linear in the quantum fields) depending on the gauge parameters 
$\x^{\ti\a}$, $\ti\a=1,\dots,n_\x$.
The gauge function, $g_a(x)$, is related to the Faddeev-Popov ghost 
operator, $(\cm c)_a(x)$, by a BRS-doublet relation:
\be
sg_a=(\cm c)_a~,~~s(\cm c)_a=0~~.\label{fpdoublet}
\ee

The gauge dependence of the classical theory, given by the action 
$\S=\S_{\rm inv}+\S_{\rm gf}$, reads\footnote{It is worthwhile to 
mention that, as pointed out by the referee, the gauge function has 
not to be necessarily linear in terms of the gauge parameters.}
\be
\x^\a{\frac{\pa\S}{\pa\x^\a}}=
s\int d^Dx~\Ld^{ab}{\ov c}_a\x^\a{\frac{\pa}{\pa\x^\a}}\left\{g_b + 
{\frac{\x^0}2} b_b\right\}~,~~\a=0,1,\dots,n_\x~~,\label{gdep}
\ee
where is assumed that no sum\footnote{When this happens will be 
indicated by the sum symbol, $\sum$.} runs on 
the gauge parameter index $\a$, {\it i.e.}, eq.(\ref{gdep}) is 
satisfied by each gauge parameter $\x^\a$ separately. Notice 
that the right-hand side of eq.(\ref{gdep}) being a BRS-variation, 
therefore, expressing the unphysical character of the gauge 
parameters, means that the physical quantities, functions of 
gauge invariant operators, are independent of those 
gauge parameters~\cite{piguet1,piguet2}.

At this point we shall introduce new Grassmann variables, 
$\c^\a$, defining them as the BRS transformations of the gauge 
parameters~\cite{sz,piguet1,piguet2}, $\x^\a$, as follows
\be
s\x^\a=\c^\a~,~~s\c^\a=0~~.\label{ebrs}
\ee
Bearing in mind the extended BRS transformations, namely, the 
fields BRS transformations (\ref{brs}) together with (\ref{ebrs}), 
we require now invariance of the classical action, $\S$, under these 
transformations, which leads to a redefinition of the gauge-fixing 
term (\ref{gf}): 
\ba
\S_{\rm gf}&=&s\int d^Dx~\Ld^{ab}{\ov c}_a\left\{g_b + 
{\frac{\x^0}2} b_b\right\}\nonumber \\
&=&\int d^Dx~\Ld^{ab}\left\{b_a [g_b + {\frac{\x^0}2}b_b] +
{\frac{\c^0}2}{\ov c}_a b_b - 
{\ov c}_a(\wt\cm c)_b\right\}~~,\label{egf}
\ea
such that
\be
sg_a=(\wt\cm c)_a~,~~s(\wt\cm c)_a=0~~.\label{efpdoublet}
\ee

It should be stressed that, in view of the dependence of the gauge 
function, $g_a(x)$, on the gauge parameters, $\x^{\ti\a}$, the requirement 
of invariance under the extended BRS transformations demands a further
redefinition of the initial Faddeev-Popov ghost operator to the new 
one, $(\wt\cm c)_a$:
\be
(\wt\cm c)_a(x)=
\int d^Dy~\left\{{\frac{\d g_a(x)}{\d \o^b(y)}}c^b(y)\right\} + 
\sum_{\ti\a=1}^{n_\x}\c^{\ti\a}{\frac{\pa g_a(x)}{\pa\x^{\ti\a}}}~~.
\ee

Since some of the BRS transformations are nonlinear, therefore, 
subjected to renormalization, an action depending on the 
antifields (external fields) coupled to those nonlinear BRS 
transformations should be added to the gauge-fixed action, $\S$:
\be
\S_{\rm ext}=\int d^Dx~\left\{A_{a_S}^{*\m}sA_\m^{a_S} + 
c_{a_S}^* sc^{a_S} + \vf^{*i}s\vf_i + 
{\bar \j}^{*I}s\j_I - s{\bar \j}^{I}\j_I^*\right\}~~,\label{ext}
\ee
where the antifields, $A_{a_S}^{*\m}$, $c_{a_S}^*$, $\vf_i^*$ and 
$\j_I^*$, are BRS invariant.

The total classical action, $\G^{(0)}$, which is the limit 
$\hbar\rightarrow 0$ of the vertex functional 
($\G$)\footnote{As we have previously mentioned, it is supposed 
{\it a priori} a renormalizable theory, meaning 
that the quantum vertex functional, $\G$, satisfies all 
symmetries of its tree-graph approximation, $\G^{(0)}$.}
\be
\G=\G^{(0)} + \co(\hbar)~~,\label{vertex}
\ee
reads
\be
\G^{(0)}=\S_{\rm inv} + \S_{\rm gf} + \S_{\rm ext}~~.\label{total}
\ee

Due to the fact that the vertex functional, $\G$, respects all symmetries 
of the classical action, $\G^{(0)}$, it does also with respect to 
the extended Slavnov-Taylor identity 
\ba
\cs(\G)&=&\int d^Dx~\left\{
{\frac{\d\G}{\d A_{a_S}^{*\m}}}{\frac{\d\G}{\d A_\m^{a_S}}} +
\pa_\m c^{a_A}{\frac{\d\G}{\d A_\m^{a_A}}} +
{\frac{\d\G}{\d c_{a_S}^*}}{\frac{\d\G}{\d c^{a_S}}} +
{\frac{\d\G}{\d \vf^{*i}}}{\frac{\d\G}{\d \vf_i}} +
{\frac{\d\G}{\d {\bar \j}^{*I}}}{\frac{\d\G}{\d \j_I}} -
{\frac{\d\G}{\d \j_I^*}}{\frac{\d\G}{\d {\bar \j}^{I}}} + 
b^a{\frac{\d\G}{\d {\ov c}^a}}\right\} + \nonumber \\
&+&\sum_{\a=0}^{n_\x}\c^\a{\frac{\pa\G}{\pa\x^\a}}= 0~~,\label{slavnov}
\ea
which expresses, in a functional way, the invariance of the quantum theory 
under the extended BRS symmetry. It is suitable to define, for later use, 
the linearized extended Slavnov-Taylor operator as bellow
\ba
\cs_\G&=&\int d^Dx~\left\{
{\frac{\d\G}{\d A_{a_S}^{*\m}}}{\frac{\d}{\d A_\m^{a_S}}} +
{\frac{\d\G}{\d A_\m^{a_S}}}{\frac{\d}{\d A_{a_S}^{*\m}}} +
\pa_\m c^{a_A}{\frac{\d}{\d A_\m^{a_A}}} +
{\frac{\d\G}{\d c_{a_S}^*}}{\frac{\d}{\d c^{a_S}}} +
{\frac{\d\G}{\d c^{a_S}}}{\frac{\d}{\d c_{a_S}^*}} +
{\frac{\d\G}{\d \vf^{*i}}}{\frac{\d}{\d \vf_i}} +
{\frac{\d\G}{\d \vf_i}}{\frac{\d}{\d \vf^{*i}}} +\right.\nonumber\\
&+&\left.{\frac{\d\G}{\d {\bar \j}^{*I}}}{\frac{\d}{\d \j_I}} +
{\frac{\d\G}{\d \j_I}}{\frac{\d}{\d {\bar \j}^{*I}}} -
{\frac{\d\G}{\d \j_I^*}}{\frac{\d}{\d {\bar \j}^{I}}} - 
{\frac{\d\G}{\d {\bar \j}^{I}}}{\frac{\d}{\d \j_I^*}} +
b^a{\frac{\d}{\d {\ov c}^a}}\right\} +
\sum_{\a=0}^{n_\x}\c^\a{\frac{\pa}{\pa\x^\a}}~~.\label{lslavnov}
\ea
The both operators, $\cs(\G)$ and $\cs_\G$, satisfy the following 
identities:
\ba
&\cs_\cf\cs(\cf)=0~,~~\forall\cf~~,&\\
&(\cs_\cf)^2=0~~{\rm if}~~\cs(\cf)=0~~,&\label{nil} 
\ea
in particular, since $\cs(\G)=0$, then a nilpotency relation 
holds for the linearized extended Slavnov-Taylor operator: 
\be
(\cs_\G)^2=0~~.\label{nilbrs}
\ee

Now, by considering the Slavnov-Taylor identity (\ref{slavnov}), 
we get 
\be
\x^\a{\frac{\pa}{\pa\c^\a}}\cs(\G)=0~~,
\ee
from which the following identities stem
\be
\x^\a{\frac{\pa\G}{\pa\x^\a}} + \c^\a{\frac{\pa\G}{\pa\c^\a}} =
\cs_\G\left(\x^\a{\frac{\pa\G}{\pa\c^\a}}\right)~~,\label{gcontrol} 
\ee
representing, therefore, identities that control the gauge dependence 
of the vertex functional $\G$. By setting now $\c^\a=0$, eq.(\ref{gcontrol}) 
reads
\be
\x^\a{\frac{\pa\G}{\pa\x^\a}} =
\cs_\G(\D_{\x^\a}\cdot\G)~~,\label{gindep} 
\ee
where $\D_{\x^\a}\cdot\G$ is the quantum operator insertion defined by 
\be
\D_{\x^\a}\cdot\G\equiv
\left.\x^\a{\frac{\pa\G}{\pa\c^\a}}\right|_{\c^\a=0}~~.
\label{opins}
\ee

As we shall see later on, the Nielsen identities~\cite{nielsen} 
are the particular case of eq.(\ref{gindep}), moreover, this equation 
reveals the unphysical character of the gauge parameters. The Nielsen
identities, which control the gauge (unphysical) dependence of the 
effective potential, stems straightforwardly from eq.(\ref{gindep}) 
when we specialize the vertex functional to the effective potential 
case. In view of the fact that the effective potential is obtained 
from the Fourier transform of the vertex functional at zero external 
momenta, $\wt\G(0,\dots,0)$, and by setting all fields to zero, except 
for the scalar fields which are taken as constant configurations, 
so, no IR problem is desired at all. This can be realized 
by guaranteeing that the gauge-fixing gives mass to 
all unphysical degrees of freedom, since the physical ones are 
assumed as massive particles in the broken phase.

From the definition of the effective potential, $\cv_{\rm eff}(\ov\vf)$, 
which is the zeroth order term in the expansion of the vertex functional, 
$\G$, by setting all fields to zero except for the scalar fields which are 
assumed as constants, $\vf_i(x)=\ov\vf_i$, 
\be
\G(\ov\vf)=- \cv_{\rm eff}(\ov\vf) \int d^Dx~~,\label{effpot1}
\ee
we get
\be
\cv_{\rm eff}(\ov\vf)\sim \sum_{n=0}^{\infty} \sum_{i_1,\dots,i_n} 
\wt\G^{i_1\cdots i_n}(0,\dots,0)~\ov\vf_{i_1}\cdots\ov\vf_{i_n}~~,
\label{effpot2}
\ee   
where $\wt\G^{i_1\cdots i_n}(0,\dots,0)$ are the momentum-space 
vertex functions taken at zero external momenta.

Now, by taking into account the gauge control identity (\ref{gindep}) 
together with the effective potential definition, eq.(\ref{effpot1}), 
the following identity holds
\be
\x^\a{\frac{\pa\cv_{\rm eff}}{\pa\x^\a}} + 
C_{\a i}(\ov\vf,\x){\frac{\pa\cv_{\rm eff}}{\pa\ov\vf_i}} = 0~~,
\label{nielsen} 
\ee 
where 
\be
C_{\a i}(\ov\vf,\x)=\left.-\int d^Dx~ 
{\frac{\d(\D_{\x^\a}\cdot\G)_\a}{\d\vf^{*i}}}\right|_{\vf_i(x)=\ov\vf_i}~~.
\ee
Notice that eq.(\ref{nielsen}) is valid for each gauge parameter $\x^\a$ 
($\a=0,\dots,n_\x$), it represents the well-known Nielsen 
identities~\cite{nielsen} for the effective potential.

As a final conclusion, this algebraic proof, based on 
general theorems of renormalization, leads to an unambiguous proof 
of the validity of the Nielsen identities, to all orders in perturbation 
theory, for renormalizable theories in any dimensions quantized in 
the context of general linear covariant gauges, with the gauge 
group being a general compact Lie group. 
The controversies presented in the literature concerning the conditional 
validity of the Nielsen identities to a subspace of the gauge parameters 
space are discussed quite detailed in ref.~\cite{our}, in fact, this letter 
is a natural extension of that work.

\underline{Acknowledgements}: 
The author expresses his gratitude to Olivier Piguet and Daniel H.T. Franco 
for helpful comments and for the critical reading of the manuscript. He 
also thanks the referee for the very pertinent comments, especially 
the one related to eq.(\ref{gdep}). He dedicates this work to his wife, 
Zilda Cristina, to his kids, Vittoria and Enzo, and to his mother, Victoria.


\begin{references}

\bibitem{nielsen} N.K.~Nielsen, Nucl.Phys. B101 (1975) 173.

\bibitem{sz} H.~Kluberg-Stern and J.-B.~Zuber, Phys.Rev. D12 (1975) 467, 
482 and 3159.

\bibitem{piguet1} O.~Piguet and K.~Sibold, Nucl.Phys. B253 (1985) 517.

\bibitem{piguet2} O.~Piguet and S.P.~Sorella, 
{\it Algebraic Renormalization}, Lecture Notes in Physics, m28, 
Springer-Verlag (Berlin-Heidelberg), 1995; see also the references 
therein.

\bibitem{our} O.M.~Del Cima, D.H.T.~Franco and O.~Piguet, hep-th/9902084, 
Nucl.Phys. B551 (1999) 813.

\bibitem{brs} C.~Becchi, A.~Rouet and R.~Stora, {\it Gauge field models} 
and {\it Renormalizable models with broken symmetries}, 
Renormalization Theory, eds. G. Velo and A.S. Wightman, D. Reidel 
(Dordrecht), 1976. 

\bibitem{bbbc} G.~Bandelloni, C.~Becchi, A.~Blasi and R.~Collina, 
Ann.Inst.Henri Poincar\'e 28 (1978) 225 and 255.

\end{references}
\end{document}